\documentclass[twocolumn,preprintnumbers,amsmath,aps]{revtex4}
\usepackage{graphicx}
\usepackage{dcolumn}
\usepackage{bm}
\usepackage{color}
\usepackage{multirow}
\usepackage{epstopdf}
\begin{document}
\newcommand{\kvec}{\mbox{{\scriptsize {\bf k}}}}
\newcommand{\qvec}{\mbox{{\scriptsize {\bf q}}}}
\def\eq#1{(\ref{#1})}
\def\fig#1{Fig.\hspace{1mm}\ref{#1}}
\def\tab#1{Tab.\hspace{1mm}\ref{#1}}
\title{Migdal-Eliashberg equations - the effective model for superconducting state in $\rm H_{3}S$}
\author{A. P. Durajski$^{\left(1\right)}$}
\email{adurajski@wip.pcz.pl}
\author{R. Szcz{\c{e}}{\'s}niak$^{\left(1, 2\right)}$}
\email{szczesni@wip.pcz.pl}
\affiliation{$^1$ Institute of Physics, Cz{\c{e}}stochowa University of Technology, Ave. Armii Krajowej 19, 42-200 Cz{\c{e}}stochowa, Poland}
\affiliation{$^2$ Institute of Physics, Jan D{\l}ugosz University in Cz{\c{e}}stochowa, Ave. Armii Krajowej 13/15, 42-200 Cz{\c{e}}stochowa, Poland}
\date{\today} 
\begin{abstract}
The high-temperature superconducting state in sulfur trihydride ($T_{C}=203$~K) has been investigated in the context of the non-adiabatic and anharmonic effects. The Migdal-Eliashberg equations and the extended Eliashberg equations, which include the lowest-order vertex corrections, have been solved numerically in the self-consistent way. For $R3m$ crystal structure, the lowest-order vertex corrections decrease the value of the Coulomb pseudopotential from $0.123$ to $0.108$. The anharmonic effects work antagonistically in relation to the vertex corrections shifting the value of $\mu^{\star}$ to $0.156$. The studies conducted for the structure $Im\overline{3}m$, where the Eliashberg function includes both the non-adiabatic and anharmonic effects, prove the even higher value of $\mu^{\star}=0.185$. Independently of the assumed method of the analysis, the nearly identical no mean-field dependence of the order parameter on the temperature was obtained: $2\Delta(0)/k_{B}T_{C}\sim 4.7$ - due to the significant strong-coupling and retardation effects: $\lambda\sim 2$ and $k_{B}T_{C}\slash \omega_{\rm \ln}\sim 0.15$-$0.19$. It means that the classical equations of Migdal-Eliashberg can be treated as a correct effective model for the superconducting state in $\rm H_{3}S$. This paper has shown that the McMillan or Allen-Dynes formulas substantially lower the value of the critical temperature in relation to the result obtained with the Eliashberg equations.\\


\end{abstract}
\maketitle
%

\section{Introduction}
The recent reports of the superconductivity in sulfur trihydride (${\rm H_{3}S}$), with the critical temperature ($T_C$) at $203$ K \cite{Drozdov2015A, Troyan2016A}, open the door to achieving the room-temperature superconductivity in the compressed hydrogen-rich materials \cite{Ashcroft2004A, Szczesniak2013D} or in the metallic hydrogen \cite{Ashcroft1968A, Cudazzo2008A, McMahon2011B, McMahon2011A, McMahon2012A}. In contrast to the cuprates \cite{Bednorz1986A, Bednorz1988A}, where the fundamental mechanism responsible for superconducting state is still debated \cite{Emery1987A, Dagotto1994A, Damascelli2003A, Bouvier2010A, Cuk2005A, Tarasewicz2006A, Szczesniak2012M, Szczesniak2014C}, the phonon-mediated pairing scenario is generally accepted in ${\rm H_{3}S}$ \cite{Mazin2016A, Bernstein2015A, Ortenzi2015A}. Therefore, in the recent theoretical papers \cite{Errea2015A, Durajski2015C, Flores-Livas2016A, Durajski2016A}, the superconducting properties of sulfur trihydride are studied in the framework of the mean-field Bardeen-Cooper-Schrieffer (BCS) theory \cite{Bardeen1957A, Bardeen1957B}, or more precisely - using the Migdal-Eliashberg (ME) approach \cite{Migdal1958A, Eliashberg1960A, Carbotte1990A}. It has been noticed that, at the ME level, it is possible to generalize the BCS model to include all many-body effects. 
In the case of the wide electron band this leads to the Eliashberg equations for the order parameter and the wave function renormalization factor \cite{Eliashberg1960A, Carbotte1990A}. From the physical point of view, the many-body effects produce the strong reduction of $T_{C}$ with respect to the standard BCS prediction, and the McMillan or Allen-Dynes formulas for the critical temperature are usually used \cite{McMillan1968A, Allen1975A}. 
The ME description of electron-phonon coupling is based on the Born-Oppenheimer or adiabatic theorem \cite{Born1927A, Born1928A}. In the considered approach, the electrons are not sensitive to the motion of the ions and are influenced only by their static electric field. The application of the ME approach to description of the electron-phonon superconductivity is justified by the very low value of the square root of the electron to the ion mass ratio $\sqrt{m/M}$ (the simplest static characteristic of the vertex corrections) \cite{Migdal1958A, Abrikosov1963A, Kirzhnits1977A}. The equivalent parameters are the ratio of velocity of sound to Fermi velocity $v_{S}/v_{F}$, the Debye energy to Fermi energy $\omega_{D}/\varepsilon_{F}$ or the ratio $\omega_{D}/W$, where $W$ is the effective band-width. Typically, the magnitude of these parameters is less than 1\% in the metals. However, even for small $\sqrt{m/M}$, there are contradictory conclusions about the importance of the vertex corrections in the normal and the superconducting state \cite{Takada1993A, Takada1995B, Takada1995A, Grimaldi1995A, Grimaldi1995B, Pietronero1995A, Danylenko1999A, Danylenko1997A, Danylenko1996A, Danylenko2001A}. In particular, there is nothing known about the meaning of the dynamic effects connected with the explicit dependence of the order parameter and the wave function renormalization factor on the frequency. Additionally, at ${\bf q}=0$, where $\bf{q}$ is the momentum transfer for the electron-phonon scattering, Migdal has already shown that the vertex corrections are not small, so they are strongly influent on the optical conductivity and Raman scattering \cite{Migdal1958A}. This was analyzed in detail by Engelsberg and Schrieffer \cite{Engelsberg1963A}. The Migdal approach is also violated for the one- and probably two-dimensional Fermi surface \cite{Allen1982A, Uemura1991A, Uemura1992A, Ambrumenil1991A}, so that the dimensionality of the examined system also plays the considerable role.
 
In the newer materials, the vertex corrections are important in the fullerene compounds \cite{Pietronero1992A, Pickett1993A}, in the cuprates \cite{Uemura1991A, Uemura1992A, Ambrumenil1991A}, in the heavy fermion systems \cite{Wojciechowski1996A}, and in the materials under high magnetic fields \cite{Goto1996A}. In particular, for $C_{60}$ compounds the following can be estimated: $\omega_{D}\sim 0.2$~eV and $\varepsilon_{F}\sim 0.2$-$0.4$~eV, so that $\omega_{D}/\varepsilon_{F}\sim 0.5$-$1$. Similarly low values of the Fermi energy ($0.1$-$0.3$~eV) are also observed in the cuprate superconductors. What is interesting, the vertex correction can be neglected in the pseudorelativistic materials (the two- and three-dimensional Dirac fermionic systems) \cite{Roy2014A}. 

The results included in the literature show that the vertex corrections can strongly affect $T_{C}$ and the other thermodynamic parameters, like the order parameter or the isotope coefficient \cite{Grimaldi1995A, Pietronero1995A, Grimaldi1995B}. This situation may be realized in the strongly correlated Fermi liquid or in the vicinity of the high peak in the density of states (the van Hove singularity \cite{Cappelluti1996A}). However, the vertex function shows the complex behavior with respect to the momentum ({\bf q}) and the frequency of the exchanged phonon. If the electron-phonon scattering is dominated by a small momentum transfer, the critical temperature and the energy gap are strong enhancement, with respect to the classical theory. The isotope effect can become negligibly small, if $\varepsilon_{F}\leq\omega_{D}$, but also anomalously large ($\alpha>1/2$) in the intermediate region. Nevertheless, Danylenko and Dolgov pay attention that the different approximations to the vertex function give quantitatively (and sometimes even qualitatively) different estimations of the self-energy \cite{Danylenko2001A}. This means that only possible accurate calculations can give the answer to the role of the vertex corrections in the normal and superconducting state. 

The importance of the vertex corrections for the system under the high pressure ($p$) on the example of ${\rm SiH_{4}}$ compound has been considered by Wei {\it et al.} \cite{Wei2010A}. It has been shown that the high phonon frequency is unfavorable to the superconductivity in the regime of the strong vertex corrections, within the assumption of the isotropic pairing in the dynamical limit. The authors concluded that the vertex corrections may even efficiently suppress $T_{C}$ approaching to the values found in the experiment ($\left[T_{C}\right]_{\rm exp}=17$~K \cite{Eremets2008A}). 

The lowest order vertex corrections for the superconducting state in ${\rm H_{3}S}$ (the crystal structure $Im\overline{3}m$) were studied by Sano {\it et al.} \cite{Sano2016A}. In the static limit with finite {\bf q}, it was shown that the vertex corrections change the critical temperature by $-34$~K ($-18$~\%). Above result is consistent with the prediction obtained for a simple model system by Grimaldi, Pietronero and Strassler \cite{Grimaldi1995A, Pietronero1995A, Grimaldi1995B}. It correlates well with the ${\rm SiH_{4}}$ results \cite{Wei2010A}. However, the calculations performed in the static limit seem to be insufficient, because the Eliashberg equations should be solved in the self-consistent way. 

It is worth to notice that the Debye phonon frequency for sulfur trihydride is approximately equal to $200$~meV and $\varepsilon_{F}\sim 16$~eV \cite{Akashi2015A}. For this reason, the ratio $\omega_{D}\slash\varepsilon_{F}$ assumes the very small value: $\sim 0.01$.  Nevertheless, Banacky in the paper \cite{Banacky2016A} suggests that the ratio $\omega_{D}\slash\varepsilon_{F}$ is approximately equal to $2\slash9$, and after including (parametrically) the H-atom displacements, the system can be found even in the anti-adiabatic regime, with: $\omega_{D}\slash\varepsilon_{F}\sim 1.6$. On the other hand, Jarlborg and Bianconi turned the attention toward the fact that the Migdal approximation might not be satisfied when the multiband electronic structure of ${\rm H_{3}S}$ would be taken into account \cite{Jarlborg2016A}. Of course, there should be an awareness that the alone value of the ratio $\omega_{D}\slash\varepsilon_{F}$ not yet finally settle the importance of the vertex corrections, because equally important are the dynamical effects connected with frequency dependence of the order parameter and the wave function renormalization factor.
 
In the case of the hydrogenated compounds, where the very high values of the critical temperature are observed \cite{Drozdov2015A, Troyan2016A, Drozdov2015B}, the values of the thermodynamic parameters should be additionally affected by the anharmonic effects. The calculations made on the anharmonic Migdal-Eliashberg level have suggested that large influence on the obtained results would be exerted by: the existence of Debye-Waller factors in the ionic potential, the change in spectral density of the one-phonon Green function induced by the phonon-phonon interactions, the many-phonon process, and the interference effects between one- and many-phonon processes \cite{Karakozov1978A}. 

With respect to the compound ${\rm H_{3}S}$ the reader should notice the results obtained by Errea {\it et al.} \cite{Errea2015A}, where it was shown that the anharmonic effects lower the value of electron-phonon coupling constant. The advanced calculations including the energy dependence of the electron density of states also confirm the significant decrease of the electron-phonon coupling constant caused by the anharmonic effects \cite{Sano2016A}. Let us notice that the influence of the anharmonic effects on the superconducting state in ${\rm H_{3}S}$ compound was not yet studied in the framework of the Eliashberg formalism that includes in the explicit way the vertex corrections to the electron-phonon interaction.
 
\vspace*{0.25 cm}
The present work studies the properties of the superconducting phase in ${\rm H_{3}S}$ characterizing with the maximum value of the critical temperature ($T_{C}=203$~K for $p=155$~GPa). Two crystal structures: $R3m$ and $Im\overline{3}m$ were taken into account. The stability of the structure $R3m$ for the pressure $155$~GPa can be shown modelling in the classical way the hydrogen nuclei (protons) \cite{Akashi2015A, Errea2016A}. However, the newest results obtained by Errea {\it et al.} \cite{Errea2016A} prove that the quantum nuclear motion is important, because it leads to the symmetrization of the length of the sulfur-hydrogen bonds, thereupon the stability is gained by the crystal structure $Im\overline{3}m$. 

Note that the literature knows the description of the properties of the superconducting state in ${\rm H_{3}S}$ for the structure $R3m$, whereas the Migdal-Eliashberg equations (MEeq) were used, as well as the spectral function calculated in the harmonic and adiabatic approximation ($\alpha^{2}F\left(\omega\right)_{{\rm H-A}}$) \cite{Durajski2016A}. In the present work, the meaning of the lowest-order vertex corrections for the crystal structure $R3m$ considered on the level of the Eliashberg equations (vEeq), in the harmonic and adiabatic approximation for the spectral function, was analyzed in the first step. Next, the influence of the anharmonism on the superconducting state was studied in the framework of the structure $R3m$ (vEeq + $\alpha^{2}F\left(\omega\right)_{{\rm AH-A}}$). The influence of the non-adiabatic effects included in the spectral function on the superconducting state can be only considered taking into account the structure $Im\overline{3}m$. Thus, the most advanced calculations were conducted for the case: vEeq + $\alpha^{2}F\left(\omega\right)_{{\rm AH-NA}}$.

\section{The Eliashberg equations with the lowest-order vertex corrections}

The set of the Eliashberg equations on the imaginary axis ($i=\sqrt{-1}$) has been taken into account \cite{Eliashberg1960A}. The Eliashberg equations were generalized to include the lowest-order vertex correction \cite{Freericks1998A}. Additionally, the momentum dependence of the electron-phonon matrix elements were neglected, which is formally equivalent to using a local approximation:
\begin{widetext}
\begin{eqnarray}
\label{r1}
\varphi_{n}&=&\pi k_{B}T\sum_{m=-M}^{M}
\frac{\lambda_{n,m}-\mu_{m}^{\star}}
{\sqrt{\omega_m^2Z^{2}_{m}+\varphi^{2}_{m}}}\varphi_{m}\\ \nonumber
&-&
\frac{\pi^{3}\left(k_{B}T\right)^{2}}{4\varepsilon_{F}}\sum_{m=-M}^{M}\sum_{m'=-M}^{M}
\frac{\lambda_{n,m}\lambda_{n,m'}}
{\sqrt{\left(\omega_m^2Z^{2}_{m}+\varphi^{2}_{m}\right)
       \left(\omega_{m'}^2Z^{2}_{m'}+\varphi^{2}_{m'}\right)
       \left(\omega_{-n+m+m'}^2Z^{2}_{-n+m+m'}+\varphi^{2}_{-n+m+m'}\right)}}\\ \nonumber
&\times&
\left[
\varphi_{m}\varphi_{m'}\varphi_{-n+m+m'}+2\varphi_{m}\omega_{m'}Z_{m'}\omega_{-n+m+m'}Z_{-n+m+m'}-\omega_{m}Z_{m}\omega_{m'}Z_{m'}
\varphi_{-n+m+m'}
\right],
\end{eqnarray}
and
\begin{eqnarray}
\label{r2}
Z_{n}&=&1+\frac{\pi k_{B}T}{\omega_{n}}\sum_{m=-M}^{M}
\frac{\lambda_{n,m}}{\sqrt{\omega_m^2Z^{2}_{m}+\varphi^{2}_{m}}}\omega_{m}Z_{m}\\ \nonumber
&-&
\frac{\pi^{3}\left(k_{B}T\right)^{2}}{4\varepsilon_{F}\omega_{n}}\sum_{m=-M}^{M}\sum_{m'=-M}^{M}
\frac{\lambda_{n,m}\lambda_{n,m'}}
{\sqrt{\left(\omega_m^2Z^{2}_{m}+\varphi^{2}_{m}\right)
       \left(\omega_{m'}^2Z^{2}_{m'}+\varphi^{2}_{m'}\right)
       \left(\omega_{-n+m+m'}^2Z^{2}_{-n+m+m'}+\varphi^{2}_{-n+m+m'}\right)}}\\ \nonumber
&\times&
\left[
\omega_{m}Z_{m}\omega_{m'}Z_{m'}\omega_{-n+m+m'}Z_{-n+m+m'}+2\omega_{m}Z_{m}\varphi_{m'}\varphi_{-n+m+m'}-\varphi_{m}\varphi_{m'}\omega_{-n+m+m'}Z_{-n+m+m'}
\right],
\end{eqnarray}
\end{widetext}
where $\varphi_{n}=\varphi\left(i\omega_{n}\right)$ represents the order parameter function and $Z_{n}= Z\left(i\omega_{n}\right)$ is the wave function renormalization factor. The Matsubara frequency is given by: $\omega_{n}=\pi k_{B}T\left(2n+1\right)$. The order parameter in the Eliashberg formalism is defined as the ratio: $\Delta_{m}=\phi_{m}/Z_{m}$ \cite{Carbotte1990A}. The electron-phonon pairing kernel can be written as:
\begin{equation}
\label{r3}
\lambda_{n,m}=2\int_0^{\omega_{D}}d\omega\frac{\omega}{\omega ^2+4\pi^{2}\left(k_{B}T\right)^{2}\left(n-m\right)^{2}}\alpha^{2}F\left(\omega\right),
\end{equation}

For the purpose of this work, the spectral function calculated by Akashi {\it et al.} \cite{Akashi2015A} in the harmonic approximation has been taken into account, where the protons are treated as the classical particles, and by Errea {\it et al.} \cite{Errea2016A} in the anharmonic case. In particular, there are two Errera's functions: first for the classical hydrogen's protons and second, which is connected with the quantum nature of the nuclei. The harmonic spectral function was obtained using the pseudopotentials method for S and H atoms implemented with the Troullier-Martin scheme \cite{Troullier1991A}. The anharmonic Eliashberg functions were calculated in the framework of the Quantum ESPRESSO code \cite{Baroni1986A, Giannozzi2009A}. The difference between the harmonic and anharmonic dynamical matrices was interpolated to the $6\times 6\times 6$ phonon momentum grid. Upon adding the harmonic matrices to the result, the anharmonic dynamical matrices were obtained. These dynamical matrices were used for the anharmonic electron-phonon coupling analysis. For the harmonic and anharmonic calculations, if the classical hydrogen's protons are considered, the $R3m$ crystal structure was assumed. Additionally, the $Im\overline{3}m$ phase was also predicted for the anharmonic case, since the quantum nature of the nucleons symmetrizes the hydrogen's bond \cite{Errea2016A}.
\begin{figure*}
\includegraphics[width=2\columnwidth]{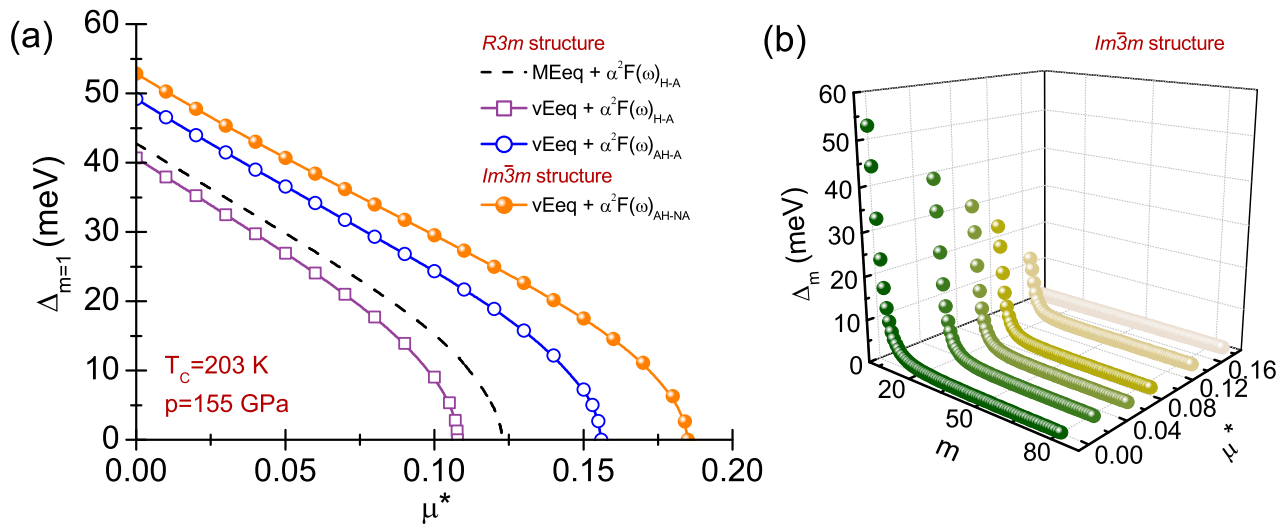}
\caption{
(a) The full dependence of the maximum value of the order parameter on the Coulomb pseudopotential at the critical temperature. The abbreviations in the figure have the following meaning: 
MEeq - the Migdal-Eliashberg equations, 
vEqe - the Eliashberg equations, which include lowest-order vertex corrections,
H - the harmonic approximation,
A - the adiabatic approximation,
AH - the anharmonic effects, and
NA - the non-adiabatic effects. 
(b) The order parameter on imaginary axis for selected values of the Coulomb pseudopotential ($T=T_{C}$).}
\label{f1}
\end{figure*}

The depairing electron correlations are modelled by: $\mu_{m}^{\star}=\mu^{\star}\theta \left(\omega_{c}-|\omega_m|\right)$, where $\mu^{\star}$ represents the Coulomb pseudopotential \cite{Morel1962A}. This term is constructed by the static screened repulsion: $\mu=\rho\left(\varepsilon_{F}\right)U$, corrected by the dynamic effects of the retarded interaction \cite{Kim1993A, Gunnarson1992A}. The symbol $\rho\left(\varepsilon_{F}\right)$ represents the density of state at the Fermi energy and $U$ is the Coulomb interaction between electrons. The Heaviside function is given by $\theta\left(x\right)$ and the cut-off frequency ($\omega_{c}$) equals ten times the Debye phonon frequency. The higher-order diagrams for the Coulomb interaction were neglected because: $\mu^{\star}-\left(\mu^{\star}\right)^{2}\simeq\mu^{\star}$.

It is worth to pay attention to the fact that the literature knows also the more general extensions of the classical Eliashberg equations than that one, which is considered in the present work. For example, Botti {\it et al.} derived the Eliashberg equations with the explicit dependence on the phonon momentum ${\bf q}$ \cite{Botti2002A}. It was shown that the order parameter removes the non-analyticity at ${\bf q}=0$ and $\omega=0$, and modifies the overall momentum structure of the vertex function. Interestingly, the strong-coupling phenomenology is naturally accounted in the non-adiabatic theory, even for the small values of the electron-phonon coupling constant. Nonetheless, due to the huge mathematical complexity of the model, the calculations could not be in the past and also cannot be presently conducted in a fully self-consistent way.

\section{Results and discussion}

The Eliashberg equations were solved using the numerical method tested in the paper \cite{Szczesniak2006B}. $2001$ Matsubara frequencies and a temperature range from $T_{0}=5$~K to $T_{C}$ were taken into account. The parameter $\mu^{\star}$ was calculated with the help of the equation: $\left[\Delta_{m=1}\left(\mu^{\star}\right)\right]_{T=T_{C}}=0$. 

The value of the Coulomb pseudopotential for the compound ${\rm H_{3}S}$ strongly depends on the accepted approach, in the framework of which the high-temperature superconducting state has been analyzed. The explicit dependence of the maximum value of the order parameter ($m=1$) on the Coulomb pseudopotential has been presented in \fig{f1}\hspace{1mm}(a). 

In the case of the structure $R3m$ (the protons from hydrogen are treated as the classical particles), in the Migdal-Eliashberg approach and for the spectral function $\alpha^{2}F\left(\omega\right)_{{\rm H-A}}$, it was obtained: $\mu^{\star}=0.123$. It is a typical value often obtained for the low-temperature superconductors with the electron-phonon pairing \cite{Carbotte1990A}. From the physical point of view it evidences of the depairing electron correlations caused by the Coulomb interaction are not being too strong in the examined system. Considering the vertex corrections of the first order on the level of the Eliashberg equations results in lowering of $\mu^{\star}$ to the value at $0.108$. This results is quite surprising, bearing in mind that the ratio $\omega_{D}\slash\varepsilon_{F}$, determining in the simplest way the static contribution to the vertex corrections, is very small ($\sim$~1\%). The noticeable lowering of the Coulomb pseudopotential’s value should be connected with the dynamical effects, which are explicitly included in the Eliashberg formalism by the dependence of the order parameter and the wave function renormalization factor on the Matsubara frequency. In the case when the anharmonic effects are also taken into account, the Coulomb pseudopotential significantly increases: $\mu^{\star}=0.156$. It means that the contributions from the vertex corrections of the first order and from the anharmonic effects are antagonistic to each other, with the advantage of the latter.

The most advanced calculations were conducted for the crystal structure $Im\overline{3}m$ (the protons from hydrogen are quantum modelled). With the help of the Eliashberg equations including the vertex corrections of the first order and the spectral function, in which the non-adiabatic and anharmonic effects are being taken into account, it has been shown that the superconducting state in ${\rm H_{3}S}$ characterizes with the high value of the Coulomb pseudopotential: $\mu^{\star}=0.185$.  However, the parameter $\mu^{\star}$ is still related to the electron depairing correlations, especially, when the influence of the retardation effects on $\mu^{\star}$ (discussed by Bauer {\it et al.} \cite{Bauer2012A}) would be taken into account.

Additionally, \fig{f1}\hspace{1mm}(b) allows to trace the full evolution of the order parameter from $\mu^{\star}$ for the structure $Im\overline{3}m$. The saturation of the order parameter for the high values of the Matsubara frequencies is clearly visible.  

\begin{figure}
\includegraphics[width=\columnwidth]{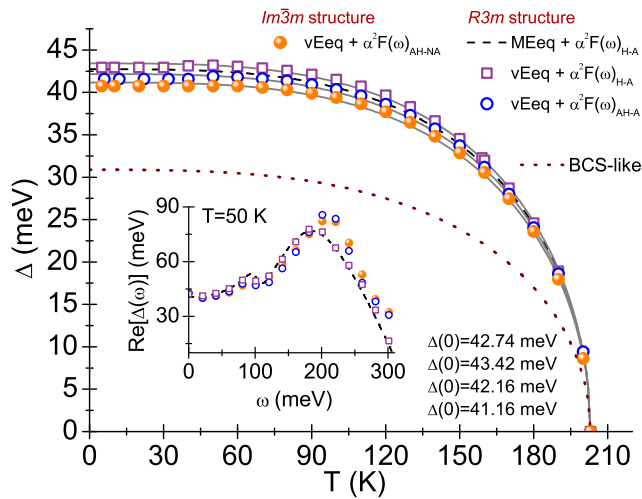}
\caption{The dependence of the order parameter on the temperature obtained for the superconductor $\rm H_{3}S$. Additionally, the shape of the function $\Delta\left(T\right)$, predicted by the BCS theory, has been plotted. The results of the numerical calculations can be reproduced with the help of the formula \eq{r4} (the grey lines). The insert presents the exemplary courses of the function ${\rm Re}\left[\Delta\left(\omega\right)\right]$ on the real axis.}
\label{f2}
\end{figure}
\begin{figure}
\includegraphics[width=\columnwidth]{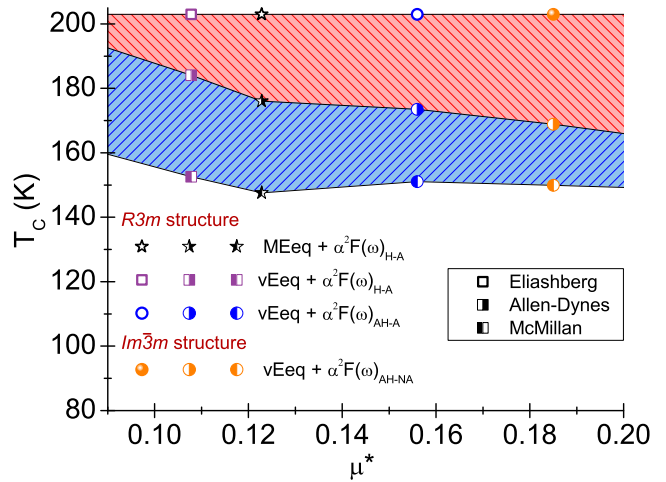}
\caption{The values of the critical temperature obtained in the framework of the Eliashberg formalism for the cases analyzed in the present work. Additionally, the results obtained with the help of the McMillan and Allen-Dynes formulas \cite{McMillan1968A, Allen1975A} have been introduced.
}
\label{f3}
\end{figure}

From the knowledge of the Coulomb pseudopotential, it is possible to obtain the temperature dependence of the order parameter. To do this, the equation \cite{Carbotte1990A}: $\Delta\left(T\right)={\rm Re}\left[\Delta\left(\omega=\Delta\left(T\right)\right)\right]$ was used. The values of the order parameter’s function on the real axis ($\omega$) were calculated from the imaginary-axis values using the analytical continuation method presented in \cite{Beach2000A}. The numerical results were shown in \fig{f2}. 

It can be noticed that the curves $\Delta\left(T\right)$, obtained in the framework of the Eliashberg formalism, very slightly differ from each other. It mean that the thermodynamic properties of the superconducting state in $\rm H_{3}S$ can be properly determined even with the help of the classical Migdal-Eliashberg equations. However, it should be boldly underlined that this is an effective model, where $\mu^{\star}$ should be treated as a fitting parameter.

Let us note that the obtained dependence $\Delta\left(T\right)$ differs very significantly from the dependence predicted by the BCS theory \cite{Bardeen1957A, Bardeen1957B}. The results of the numerical calculations can be reproduced with the help of the formula:
\begin{equation}
\label{r4} 
\Delta\left(T\right)\slash\Delta\left(0\right)=\sqrt{1-\left(T\slash T_{C}\right)^{\Gamma}},
\end{equation}
where the temperature exponent $\Gamma$ is equal to $3.26$. Note that the BCS theory predicts: $\Gamma=3$ \cite{Eschrig2001A}. Additionally, in the framework of the BCS model the ratio $2\Delta(0)/k_{B}T_{C}$ is equal to $3.53$. In the case of sulfur trihydride, it was obtained: $2\Delta(0)/k_{B}T_{C}=4.7$ (the structure $Im\overline{3}m$). The result above means that the compound $\rm H_{3}S$ should be included in the group of the superconductors being in the strong-coupling area. This area is determined by two conditions: (i) high value of the electron-phonon coupling constant: $\lambda=2\int^{\omega_{D}}_0 d\omega\alpha^2\left(\omega\right)F\left(\omega\right)\slash\omega$, and (ii) significant retardation and many-body effects. For the superconductor $\rm H_{3}S$ the electron-phonon coupling constant is approximately equal $2$ (precisely $1.94$ for the structure $Im\overline{3}m$ \cite{Errea2016A}). The retardation and many-body effects can be on the other hand characterized with the help of the parameter $k_{B}T_{C}\slash\omega_{\rm ln}$ \cite{Carbotte1990A}, which for the crystal structure $R3m$ takes the values from the range from $0.164$ to $0.192$, while for $Im\overline{3}m$ we obtain $0.153$. In the BCS limit the Eliashberg equations give: $k_{B}T_{C}\slash\omega_{\rm ln}\rightarrow 0$. The quantity $\omega_{{\rm ln}}$ is called the phonon logarithmic frequency: $\omega_{{\rm ln}}=\exp\left[\frac{2}{\lambda}\int^{\omega_{D}}_{0}d\omega\alpha^{2}\left(\omega\right)F\left(\omega\right) \ln\left(\omega\right)\slash\omega\right]$.

The high real value of the Coulomb pseudopotential ($\mu^{\star}=0.185$) for $\rm H_{3}S$ superconductor has also one significant consequence. Namely, for the calculation of the critical temperature’s value it is not relevant to use the approximate McMillan and Allen-Dynes formulas \cite{McMillan1968A, Allen1975A} (see Appendix A), because they allow to obtain relatively correct $T_{C}$ only for the low values of $\mu^{\star}$. The illustration of the above statement are the results presented in \fig{f3}, where the values of the critical temperature, obtained in the framework of the Eliashberg formalism and with the help of the analytical formulas, have been plotted. It can be easily noticed that together with the increasing Coulomb pseudopotential, the analytical results (Allen-Dynes) become more and more inaccurate. For example, for $\mu^{\star}=0.185$, the McMillan formula underestimates the value of the critical temperature at $26$~\%, and the Allen-Dynes formula at $17$~\%. 

\begin{figure}
\includegraphics[width=\columnwidth]{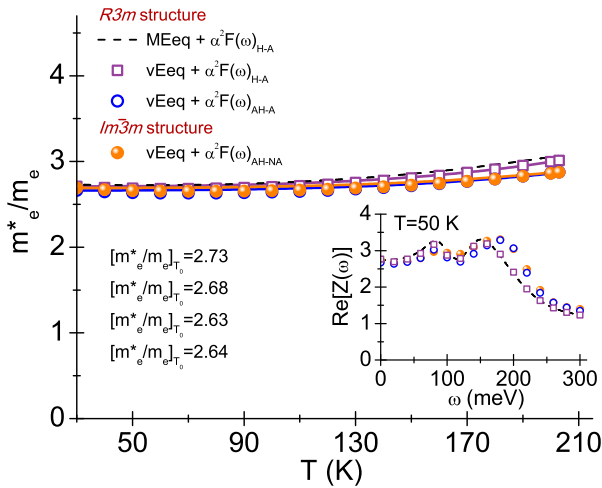}
\caption{The ratio of the electron effective mass to the electron band mass as a function of the temperature for $\rm H_{3}S$ superconductor. The insert presents the exemplary courses of the function ${\rm Re}\left[Z\left(\omega\right)\right]$ on the real axis.}
\label{f4}
\end{figure}

The electron effective mass ($m^{\star}_{e}$) has been calculated in the last step from the relation: 
$m^{\star}_{e}={\rm Re}\left[Z\left(\omega=0\right)\right]m_{e}$, where $m_{e}$ denotes the electron band mass. The dependence of the ratio $m^{\star}_{e}\slash m_{e}$ on the temperature has been plotted in \fig{f4}. Additionally, the insert presents the function ${\rm Re}\left[Z\left(\omega\right)\right]$ for the exemplary temperature at $50$~K. 

The obtained results prove that the electron effective mass is large ($m^{\star}_{e}\sim 3m_{e}$) in the whole range of the temperature - from $T_{0}$ to the critical temperature. It reaches its maximum value $m^{\star}_{e}$ always at $T_{C}$. This feature is characteristic for the wave function renormalization factor, and is connected with the vanishing order parameter. In particular, for the crystal structure $Im\overline{3}m$, where $\mu^{\star}=0.185$, it was obtained: $\left[m^{\star}_{e}\slash m_{e}\right]_{\rm max}=2.87$. 

\section{Conclusions}

The properties of the high-temperature superconducting state in $\rm H_{3}S$ compound have analyzed in the strong-coupling formalism ($T_{C}=203$~K). The classical Migdal-Eliashberg equations and the Eliashberg equations, which included the lowest-order vertex corrections, were taken into account. The equations have been solved in the self-consistent way.

In the case when protons from hydrogen are described as the classical particles, stable is the crystal structure $R3m$ \cite{Akashi2015A, Errea2016A}. It has been shown that the non-adiabatic effects related to the lowest-order vertex corrections lower the value of the Coulomb pseudopotential from $0.123$ to $0.108$. The obtained result is connected with the explicit dependence of the order parameter and the wave function renormalization factor on the Matsubara frequency. Let us note that the analogous result was obtained for ${\rm SiH_{4}}$ compound located under the influence of the high pressure \cite{Wei2010A}. For compound $\rm H_{3}S$ the more important than the lowest-order vertex corrections are the anharmonic effects. They eliminate entirely the influence of the vertex corrections on the properties of the superconducting state and they contribute to a significant increase in the value of the Coulomb pseudopotential: $\mu^{\star}=0.156$.

The most advanced calculations were performed for the crystal structure $Im\overline{3}m$, which turns out to be stable in the case, when the protons from hydrogen are described in terms of the quantum \cite{Errea2016A}. The numerical results obtained with the help of the Eliashberg equations for the spectral function including the non-adiabatic and anharmonic effects prove that the value of the Coulomb pseudopotential increases up to $0.185$. From the physical points of view, it is the quite high value. However, $\mu^{\star}$ is still related to the electron depairing correlations. 

Independently of the approach, in the framework of which the superconducting state in $\rm H_{3}S$ compound was analyzed, we have obtained almost identical dependence of the order parameter on the temperature. It means that the thermodynamic properties of the superconducting state can be properly determined even in the framework of the classical Migdal-Eliashberg formalism, as long as the effective value of the Coulomb pseudopotential is correctly determined (from the equation: $\left[\Delta_{m=1}\left(\mu^{\star}\right)\right]_{T=T_{C}}=0$). It should be emphasized that the shape of the function $\Delta\left(T\right)$ significantly deviates from the shape predicted by the BCS theory. In particular, the ratio $2\Delta(0)/k_{B}T_{C}$ is equal to $4.7$, and the temperature exponent $\Gamma$ equals $3.26$. The result above is related to the significant strong-coupling and retardation effects present in $\rm H_{3}S$. Those effects contribute also to the high value of the electron effective mass: $m^{\star}_{e}\sim 3 m_{e}$. 

The high physical value of the Coulomb pseudopotential ($\mu^{\star}=0.185$) means that the critical temperature cannot be properly calculated with the help of the McMillan and the Allen-Dynes formulas. The obtained inaccuracies of the analytical formulas are equal respectively to $26$~\% and $17$~\%.
 
\section{Acknowledgments}

Authors are grateful to Marcin Mierzejewski (University of Silesia) and Ryosuke Akashi (University of Tokyo) for scientific discussion.

\appendix
\section{\label{Appendix A} The characteristics of the electron-phonon interaction in ${\rm H_{3}S}$ and the McMillan and Allen-Dynes formulas}

\tab{t1} collects the parameters characterizing the electron-phonon interaction in ${\rm H_{3}S}$ compound. Those quantities can serve inter alia to calculate the values of the critical temperature with the help of the McMillan \cite{McMillan1968A} or Allen-Dynes \cite{Allen1975A} formulas. In particular, the McMillan formula can be written as follows:
\begin{equation}
\label{rA1}
k_{B}T_{C}=\frac{\omega_{\ln}}{1.2}\exp\left[\frac{-1.04\left(1+\lambda\right)}
{\lambda-\mu^{\star}\left(1+0.62\lambda\right)}\right].
\end{equation}
\begin{table}
\caption{\label{t1} The values of the selected parameters characterizing the electron-phonon interaction in compound ${\rm H_{3}S}$. The results were obtained on the basis of the spectral functions calculated in the papers \cite{Akashi2015A} and \cite{Errea2016A}. 
In addition: $\omega_{2}=\frac{2}{\lambda}\int_{0}^{\omega_{D}}d\omega\alpha^{2}F\left(\omega\right)\omega$.}
\begin{ruledtabular}
\begin{tabular}{cccc}
Structure                          & $R3m$                  & $R3m$            & $Im\overline{3}m$      \\ 
                                   &                        &                  &                        \\
$\alpha^{2}F\left(\omega\right)$   & $H-A$	                & $AH-A$	       & $AH-NA$                \\ \hline
                                   &                        &                  &                        \\
$\omega_{D}$ (meV)                 & 207.25                 & 224.69           & 231.18                 \\
                                   &                        &                  &                        \\
$\lambda$                          & 2.07	                & 1.93	           & 1.94                   \\
                                   &                        &                  &                        \\
$\omega_{\rm ln}$ (meV)            & 90.95                  & 106.44	       & 114.56                 \\
                                   &                        &                  &                        \\
$\sqrt{\omega_{2}}$ (meV)          & 117.75                 & 134.52           & 137.54                 \\
\end{tabular}
\end{ruledtabular}
\end{table}
The more accurate Allen-Dynes formula has a following form:
\begin{equation}
\label{rA2}
k_{B}T_{C}=f_{1}f_{2}\frac{\omega_{{\rm ln}}}{1.2}\exp\left[\frac{-1.04\left(1+\lambda\right)}
{\lambda-\mu^{\star}\left(1+0.62\lambda\right)}\right],
\end{equation}
where $f_{1}$ and $f_{2}$ are the correction functions:
\begin{equation}
\label{rA3}
f_{1}=\left[1+\left(\frac{\lambda}{\Lambda_{1}}\right)^{\frac{3}{2}}\right]^{\frac{1}{3}},
\end{equation}
\begin{equation}
\label{rA4}
f_{2}=1+\frac{\left(\frac{\sqrt{\omega_{2}}}{\omega_{\rm{ln}}}-1\right)\lambda^{2}}
{\lambda^{2}+\Lambda^{2}_{2}}.
\end{equation}
Additional marks are introduced:
\begin{equation}
\label{rA5}
\Lambda_{1}=2.46(1+3.8\mu^{\star}),
\end{equation}
and 
\begin{equation}
\label{rA6}
\Lambda_{2}=1.82(1+6.3\mu^{\star})\frac{\sqrt{\omega_{2}}}{\omega_{\ln}}.
\end{equation}
%

\bibliography{Bibliography}
\end{document}